Effect of time of day on reward circuitry:

Further thoughts on methods, prompted by Steel et al 2018.

Greg Murray[*,1], Catherine Orr[1], Jamie E. M. Byrne[1], Matthew E. Hughes[1,4], Susan L. Rossell[1,3] and Sheri L. Johnson[5]

Swinburne University of Technology

Author note

[1] Centre for Mental Health, Faculty Health, Arts and Design, Swinburne University, Hawthorn, VIC, Australia

[3] Psychiatry, St Vincent's Hospital, Fitzroy, Melbourne, VIC, Australia

[4] The Australian National Imaging Facility

[5] University of California, Berkeley, California, USA

* Correspondence concerning this article should be addressed to Professor Greg Murray, Centre for Mental Health, Faculty Health, Arts and Design, Swinburne University of Technology, John St, Hawthorn VIC, Australia, 3122. Email: gwm@swin.edu.au

# Circadian modulation of reward using fMRI: Overcoming methodological challenges

**Abstract**

The interplay between circadian and reward function is well understood in animal models, and is of growing interest as an aetiological explanation in psychopathologies. Circadian modulation of reward function has been demonstrated in human behavioural data, but understanding at the neural level is limited. In 2017, our group published results of a first step in addressing this deficit, demonstrating a diurnal rhythm in fMRI-measured reward activation. In 2018, Steel et al wrote a constructive critique of our findings, and the aim of this paper is to outline how future research could improve on our first proof-of-concept study. Key challenges include addressing divergent and convergent validity (by addressing non-reward neural variation, and testing for absence of variation in threat-related pathways), preregistration and power analysis to protect against false positives, wider range of fMRI methods (to directly test our post-hoc hypothesis of some form of reward prediction error, and multiple phases of reward), the parallel collection of behavioural data (particularly self-reported positive affect, and actigraphically measured activity) to illuminate the nature of the reward activation across the day, and some attempt to parse out circadian versus homeostatic/masking influences on any observed diurnal rhythm in neural reward activation.

Understanding the adaptive interplay between biobehavioural systems has potential to advance understanding of motivated behaviour. Our group has conducted a series of studies into one particular relationship, viz. the possibility that human reward functioning is modulated by circadian function (e.g., Murray et al., 2009, see Figure 1). There are multiple reasons to think that this interplay is important in both normal functioning and multiple psychopathologies (e.g., Nusslock and Alloy, 2017, Alloy et al., 2015, Webb, 2017). Our recent publication (Byrne et al., 2017) reported on a preliminary study investigating this relationship at the level of reward neurobiology. In a letter to Journal of Neuroscience, Steel et al (Steel et al., 2018) provided a reasonable and constructive critique of our study, and here we seek to encourage further research into this question by engaging with their criticisms.

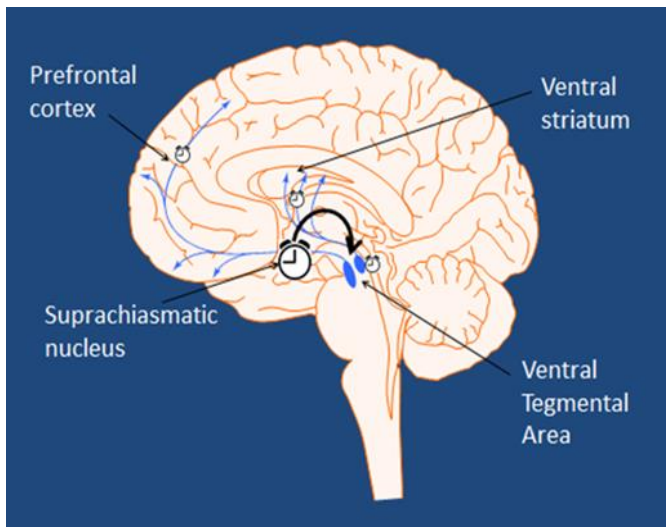

Figure 1: Putative pathways by which circadian function may moderate reward in humans (inferred from animal data). Shown are indirect pathways from SCN to VTA (black arrow) and gene expression in reward centres (small clock icons). Ascending dopaminergic pathways shown in blue.

*1. What did we find?*
Using a novel three time-point repeated measures fMRI design, we found diurnal variation in one reward region (specifically, the left putamen) to a validated and commonly-used reward task in healthy young males. We concluded that, within the study's limitations, this provisional finding was consistent with predicted circadian modulation of reward.

Based on an evolutionary hypothesis of reward priming in the brain (Clark et al., 1989), and replicated findings about activation patterns in the putamen (Schultz, 2016), we tentatively interpreted the particular waveform observed (left putamen activation at its nadir in the early afternoon compared with morning and evening) as evidence of a form of reward prediction error.

*2. Controls for multiple testing*
Our a priori regions of interest (ROI) were mPFC, VTA, anterior cingulate cortex, caudate, NAc and putamen as defined by the Automated Anatomical Labelling (AAL) atlas (Tzourio-Mazoyer et al., 2002). The primary analysis involved modelling the whole brain data, followed by a small volume correction (SVC) analysis for each ROI in isolation. We used an uncorrected voxel testing with *p* < .001 (which was not a liberal standard at the time of those analyses). We reported that 23 voxels in the left putamen were found in the SVC analysis. The cluster corrected significance for the left putamen alone was *p* = .014, and *p* = .055 when bilateral putamen and bilateral caudate nuclei were combined in a single mask. We agree with Steel et al's suggestion that our failure to control for the testing of all regions within the mask has advantaged our hypothesis, risking a false positive finding.

Steel et al. note that we also reported some results not surviving correction. These analyses were included at the request of the paper's reviewers: Given the exploratory nature of this work, reviewers suggested investigating the contrasts for Reward > Loss. Steel et al. are correct that we did not conduct an analysis showing the diurnal effect was specific to the left putamen, but we did not claim that activation was specific to the left putamen. Our prediction was that a diurnal rhythm would be found in one or more reward regions in response to monetary reward task, and future research should investigate whether such specificity exists.

*Reward variation should be interrogated in the context of diurnal variation in biology*
We concur with Steel et al that complex interactions between time of day, physiology, and measurement techniques make interpretation of our findings challenging. Recent evidence for a holistic circadian variation in brain function (Muto, 2016, Ly et al., 2016, Chellappa et al., 2016) underscores the importance of exploring specificity of any observed diurnal variation in reward-related regions.

As suggested by Steel et al, tracking variables such as salivary cortisol across the day would be consistent with the aim of distinguishing reward-specific processes from the cyclic background of the nervous system more broadly. We do not agree that simply measuring variables that also have a circadian rhythm (e.g., melatonin and cortisol) would strengthen the argument for a *circadian provenance* of the observed diurnal rhythm in putamen activation.

*3. Findings may not generalise from the particular stimulus used here*
Steel et al rightly point out that our finding is limited to one task (Human Connectome Project gambling task), and one stimulus type. While there is a common brain network that has been shown to activate across different reward stimuli (see, Sescousse et al., 2013), we agree with Steel et al that generalisation to other stimuli dimensions would be an improved design. One of the small set of repeated measures studies relevant to this question, for example, used food stimuli to generate complex relationship between subjective appetite reports and neural reward activation patterns across the day (Masterson et al., 2016).

*Without measures of reward behaviour, the neural effect is difficult to interpret*
We agree it is important to relate the observed neural changes to behavioural measures of reward activation, that is, we need to demonstrate that the observed neural variation is behaviourally meaningful. Through experimenter error, our planned collection of positive affect data alongside fMRI sessions did not occur. There is growing interest in multi-modal investigation of human reward processing. Wang and colleagues (2016), for example, have argued for meaningful measurement of behavioural traits such as reward sensitivity and impulsivity as part of a systematic approach to investigating the behavioural components of reward functioning in the context of fMRI studies.

*4. Our interpretation of the derived diurnal pattern as consistent with a reward prediction error needs to be directly tested*

We argued that reward prediction error provides one possible explanation of the specific pattern of diurnal variation observed. Specifically, we suggested that reward activation being at a relative nadir around early afternoon could reflect the brain being primed (theoretically by the circadian system) to expect rewards at this time, compared to morning and evening time points. Given that the putamen responds to unexpected rewards, this 'pleasant surprise' hypothesis posits that circadian priming to expect rewards around early afternoon appears as relatively increased putamen activation at 1000 and 1900 hours compared with 1400 hours.

How does this interpretation sit with replicated evidence that self-reported positive affect *peaks* in the early afternoon (e.g., Murray et al., 2002)? There are two understandings of the relationship between positive affect and biobehavioural reward motivation. One approach posits that positive affect is the subjective manifestation of reward activation (e.g., Watson et al., 1999, Knutson et al., 2014); under this viewpoint, our finding of a nadir in neural reward activation at 1400 hours is inconsistent with the data showing positive affect is at its zenith at this time. The alternate control process model (Carver and Scheier, 1990), posits that positive affect is a barometer not of reward activation but of the organism's successful progress towards reward. Under this view, it is entirely possible that putamen response to reward at 1400 is relatively decreased under fMRI conditions, while under naturalistic conditions positive affect is maximal at the same time. Inferences about this issue are made more complex by, (a) different reward regions potentially having different relationships to positive affect, (b) differences between 1400 in the scanner with artificial reward stimuli versus 1400 in the real world with an ecologically-valid array of opportunities and threats, (c) differences between the circadian and the non-circadian components of rhythms in both neural activation and positive affect.

*5. Where to next?*
Steel et al generously describe our question as new and interesting, and encourage its investigation using more rigorous methods. Emerging from the above considerations, we propose nine issues for a future series of studies to consider.

(a) There is growing recognition in fMRI research that underpowered studies not only increase Type II but also Type I error. Future research in this area should use a formal statistical power analysis to calculate required sample sizes. To avoid well-recognised problems with multiple suites of analytic steps (Poldrack and colleagues speculate that there are more recognised approaches to fMRI analysis, than there are fMRI publications, Poldrack et al., 2017), studies must be formally pre-registered, and describe inclusion and exclusion criteria, software workflows (including contrasts and multiple-comparison methods), and operational definitions for all planned regions of interest.

(b) Future research should also attend to both convergent and divergent validity. Convergent validity will improve by measuring reward system function at multiple levels beyond the neural. Obvious candidate constructs are self-reported positive affect (e.g., Miller et al., 2015), and locomotor activity levels across the day (e.g., Lyall et al., 2018). To improve divergent validity, designs should also test for diurnal variation in processes (e.g. threat activation) which are postulated to not exhibit such variation (Murray et al., 2009). The specificity of the hypothesised diurnal variation in reward activation should also be tested by systematically accounting for non-reward related variation in brain function across the day.

(c) Future studies should seek to directly test our 'reward prediction error' explanation of the pattern of diurnal variation observed. The Monetary Incentive Delay Task (Knutson et al., 2000) is an event-related fMRI task that temporally separates the reward cue and reward receipt, permitting calculation of the reward prediction error signal.

(d) Even when the time-series variable is simple behavioural measures of higher cortical functions, time-of-day effects can be parsed into homeostatic, sleep proximity, and circadian components. When the time-series variable is fMRI-measured patterns of brain activation, the pragmatic challenges are magnified. Ideally, future research should employ more than three time points across the day, to address the post-lunch dip that occurs in some activation-related variables in some individuals, and the dynamics of the wake maintenance zone. The greater precision of more time points needs to be weighed against participant burden, and repeated testing effects. Based on our research in diurnal rhythms in self-reported variables, we propose that four time points may provide the optimal balance of precision versus burden.

(e) To address the individual difference confounds of chronotype and typical sleep phase, one design strategy for future studies like Byrne et al would be to split the sample into early versus late chronotypes. This distinction is theoretically meaningful because later chronotype has been reliably associated with risk of psychopathology, while early chronotypes are less likely to have problems with sleep, mood or substance use (e.g., Ottoni et al., 2012, Melo et al., 2016, DeYoung et al., 2007). The two-group approach would also have methodological advantages, permitting four time points reasonably spaced across the day to optimise sensitivity to a diurnal rhythm without disturbing participants' normal sleep. The Early chronotype group could be tested at 0900, 1200, 1500, 1800 without curtailing the normal sleep cycle; similarly the Late chronotype group could be tested at 1200, 1500, 1800 and 2100. Obviously this would be a resource-intensive study design, as both groups would need to be adequately powered.

(f) To our knowledge, there has been little consideration in fMRI literature of how to operationalise predictions about the interaction between an endogenously primed brain state and an exogenous stimulus. Our underlying model is that the brain is primed for reward-related behaviour in a diurnal rhythm - whether this ought to appear as relatively increased or decreased reward-centre activation in a given reward centre *to a presented reward stimulus* is unclear. It is possible that the most sensitive measure of our prediction would eschew the complex interaction implied by the combination of temporal priming and administered reward stimuli. Resting state analyses may provide an important perspective on the neural priming that we are hypothesising (Goel et al., 2013), and open up this question to investigation from a functional integration viewpoint (Wang et al., 2016). To our knowledge, no study has systematically investigated diurnal variation in reward-related components of resting state networks, but one study generated some support (Blautzik et al., 2013).

(g) It may also be important to distinguish different phases of the reward process, under Berridge's tripartite wanting, liking, learning model (Berridge and Kringelbach, 2008). While our research has emphasised the motivational/wanting facet of the reward process (Byrne and Murray, 2017), animal research has generated evidence for circadian modulation of liking and learning facets, and a recently published study found complex evidence of diurnal variation in reward learning in humans (Whitton et al., 2018). There is growing interest in parsing these components of reward in the context of fMRI (Wang et al., 2016).

(h) It is also important to note that the particular circadian X reward interaction tested here (the Circadian Reward Rhythm, Murray et al., 2017) is only one of numerous possible types of interaction. At this whole-person level, for example, moderation of circadian function by reward-driven activation in the environment (think the person with bipolar disorder disturbing their light exposure by working late into the night) is an important reciprocal interaction.

(i) Finally, the repeated measures approach used here is not the only way to progress the question of interest. A useful parallel tack to the current question would be to move from a within- to a between-subject focus. This would support researchers exploiting publicly available data (e.g., 1000 Functional Connectomes Project/International Neuroimaging Data-sharing Initiative, the Consortium for Reliability and Reproducibility, OpenfMRI, Human Connectome Project) on the relation between reward activation and time of day in the.

*6. Conclusions*

Like the field of genetics before it, fMRI research has now embraced the problem of methodological divergence across studies, and particularly the problem of false positive findings (Poldrack et al., 2017). We are comfortable that Byrne et al (2017) has not misdirected the literature in this burgeoning area, because the paper describes the study's aim as exploratory, and conclusions were described as preliminary and provisional. Nonetheless, we appreciate the commentary provided by Steel et al (2018). We consider the core finding of Byrne et al as a proof of concept, viz. repeated measures of task-based activation in reward centres can be used to identify a signal of diurnal variation at the group level. We are confident that future investigation of this important interaction hypothesis is warranted, and Steel et al have helped identify important ways in which future more rigorous investigations can generate more convincing findings. We are also confident that the particular finding of a diurnal rhythm in left putamen activation *will not be* the final word in this area.

**References**

ALLOY, L. B., NUSSLOCK, R. & BOLAND, E. M. 2015. The development and course of bipolar spectrum disorders: an integrated reward and circadian rhythm dysregulation model. *Annu Rev Clin Psychol,* 11**,** 213-50.

BERRIDGE, K. C. & KRINGELBACH, M. L. 2008. Affective neuroscience of pleasure: Reward in humans and animals. *Psychopharmacology,* 199**,** 457-480.

BLAUTZIK, J., VETTER, C., PERES, I., GUTYRCHIK, E., KEESER, D., BERMAN, A., KIRSCH, V., MUELLER, S., POPPEL, E., REISER, M., ROENNEBERG, T. & MEINDL, T. 2013. Classifying fMRI-derived resting-state connectivity patterns according to their daily rhythmicity. *Neuroimage,* 71**,** 298-306.

BYRNE, J. E. & MURRAY, G. 2017. Diurnal rhythms in psychological reward functioning in healthy young men: 'Wanting', liking, and learning. *Chronobiol Int,* 34**,** 287-295.

BYRNE, J. E. M., HUGHES, M. E., ROSSELL, S. L., JOHNSON, S. L. & MURRAY, G. 2017. Time of Day Differences in Neural Reward Functioning in Healthy Young Men. *J Neurosci,* 37**,** 8895-8900.

CARVER, C. S. & SCHEIER, M. F. 1990. Origins and Functions of Positive and Negative Affect - a Control-Process View. *Psychological Review,* 97**,** 19-35.

CHELLAPPA, S. L., GAGGIONI, G., LY, J. Q. M., PAPACHILLEOS, S., BORSU, C., BRZOZOWSKI, A., ROSANOVA, M., SARASSO, S., LUXEN, A., MIDDLETON, B., ARCHER, S. N., DIJK, D. J., MASSIMINI, M., MAQUET, P., PHILLIPS, C., MORAN, R. J. & VANDEWALLE, G. 2016. Circadian dynamics in measures of cortical excitation and inhibition balance. *Scientific Reports,* 6.

CLARK, L. A., WATSON, D. & LEEKA, J. 1989. Diurnal variation in the positive affects. *Motivation and Emotion,* 13**,** 205-234.

DEYOUNG, C. G., HASHER, L., DJIKIC, M., CRIGER, B. & PETERSON, J. B. 2007. Morning people are stable people: Circadian rhythm and the higher-order factors of the Big Five. *Personality and Individual Differences,* 43**,** 267-276.

GOEL, N., BASNER, M., RAO, H. & DINGES, D. F. 2013. Circadian rhythms, sleep deprivation, and human performance. *Progress in molecular biology and translational science,* 119**,** 155.

KNUTSON, B., KATOVICH, K. & SURI, G. 2014. Inferring affect from fMRI data. *Trends in Cognitive Sciences,* 18**,** 422-428.

KNUTSON, B., WESTDORP, A., KAISER, E. & HOMMER, D. 2000. FMRI visualization of brain activity during a monetary incentive delay task. *Neuroimage,* 12**,** 20-27.

LY, J. Q. M., GAGGIONI, G., CHELLAPPA, S. L., PAPACHILLEOS, S., BRZOZOWSKI, A., BORSU, C., ROSANOVA, M., SARASSO, S., MIDDLETON, B., LUXEN, A., ARCHER, S. N., PHILLIPS, C., DIJK, D. J., MAQUET, P., MASSIMINI, M. & VANDEWALLE, G. 2016. Circadian regulation of human cortical excitability. *Nature Communications,* 7.

LYALL, L. M., WYSE, C. A., GRAHAM, N., FERGUSON, A., LYALL, D. M., CULLEN, B., CELIS MORALES, C. A., BIELLO, S. M., MACKAY, D., WARD, J., STRAWBRIDGE, R. J., GILL, J. M. R., BAILEY, M. E. S., PELL, J. P. & SMITH, D. J. 2018. Association of disrupted circadian rhythmicity with mood disorders, subjective wellbeing, and cognitive function: a cross-sectional study of 91 105 participants from the UK Biobank. *The Lancet Psychiatry,* 5**,** 507-514.

MASTERSON, T. D., KIRWAN, C. B., DAVIDSON, L. E. & LECHEMINANT, J. D. 2016. Neural reactivity to visual food stimuli is reduced in some areas of the brain during evening hours compared to morning hours: an fMRI study in women. *Brain Imaging Behav,* 10**,** 68-78.

MELO, M. C. A., ABREU, R. L. C., LINHARES NETO, V. B., DE BRUIN, P. F. C. & DE BRUIN, V. M. S. 2016. Chronotype and circadian rhythm in bipolar disorder: A systematic review. *Sleep Medicine Reviews.*

MILLER, M. A., ROTHENBERGER, S. D., HASLER, B. P., DONOFRY, S. D., WONG, P. M., MANUCK, S. B., KAMARCK, T. W. & ROECKLEIN, K. A. 2015. Chronotype predicts positive affect rhythms measured by ecological momentary assessment. *Chronobiol Int,* 32**,** 376-84.

MURRAY, G., ALLEN, N. B. & TRINDER, J. 2002. Mood and the circadian system: Investigation of a circadian component in positive affect. *Chronobiology International,* 19**,** 1151-1169.

MURRAY, G., ALLOY, L. B. & NUSSLOCK, R. Reward and circadian interactions in bipolar disorder: models, mechanisms, clinical implications.  International Society for Bipolar Disorders 19th Annual Conference, 2017 Washingon.

MURRAY, G., NICHOLAS, C. L., KLEIMAN, J., DWYER, R., CARRINGTON, M. J., ALLEN, N. B. & TRINDER, J. 2009. Nature's clocks and human mood: the circadian system modulates reward motivation. *Emotion,* 9**,** 705-16.

MUTO, V. 2016. Local modulation of human brain responses by circadian rhythmicity and sleep debt (vol 354, aam5837, 2016). *Science,* 354**,** 1543-1543.

NUSSLOCK, R. & ALLOY, L. B. 2017. Reward processing and mood-related symptoms: An RDoC and translational neuroscience perspective. *J Affect Disord,* 216**,** 3-16.

OTTONI, G. L., ANTONIOLLI, E. & LARA, D. R. 2012. Circadian preference is associated with emotional and affective temperaments. *Chronobiol Int,* 29**,** 786-93.

POLDRACK, R. A., BAKER, C. I., DURNEZ, J., GORGOLEWSKI, K. J., MATTHEWS, P. M., MUNAFO, M. R., NICHOLS, T. E., POLINE, J. B., VUL, E. & YARKONI, T. 2017. Scanning the horizon: towards transparent and reproducible neuroimaging research. *Nature Reviews Neuroscience,* 18**,** 115-126.

SCHULTZ, W. 2016. Dopamine reward prediction error coding. *Dialogues in Clinical Neuroscience,* 18**,** 23-32.

SESCOUSSE, G., CALDU, X., SEGURA, B. & DREHER, J. C. 2013. Processing of primary and secondary rewards: A quantitative meta-analysis and review of human functional neuroimaging studies. *Neuroscience and Biobehavioral Reviews,* 37**,** 681-696.

STEEL, A., THOMAS, C. & BAKER, C. I. 2018. *Effect of time of day on reward circuitry. A discussion of Byrne et al. 2017* [Online].  [Accessed].

TZOURIO-MAZOYER, N., LANDEAU, B., PAPATHANASSIOU, D., CRIVELLO, F., ETARD, O., DELCROIX, N., MAZOYER, B. & JOLIOT, M. 2002. Automated anatomical labeling of activations in SPM using a macroscopic anatomical parcellation of the MNI MRI single-subject brain. *Neuroimage,* 15**,** 273-289.

WANG, K. S., SMITH, D. V. & DELGADO, M. R. 2016. Using fMRI to study reward processing in humans: past, present, and future. *Journal of Neurophysiology,* 115**,** 1664-1678.

WATSON, D., WIESE, D., VAIDYA, J. & TELLEGEN, A. 1999. The two general activation systems of affect: Structural findings, evolutionary considerations, and psychobiological evidence. *Journal of Personality and Social Psychology,* 76**,** 820-838.

WEBB, I. C. 2017. Circadian Rhythms and Substance Abuse: Chronobiological Considerations for the Treatment of Addiction. *Curr Psychiatry Rep,* 19**,** 12.

WHITTON, A. E., MEHTA, M., IRONSIDE, M. L., MURRAY, G. & PIZZAGALLI, D. A. 2018. Evidence of a diurnal rhythm in implicit reward learning. *Chronobiol Int,* 35**,** 1104-1114.